\documentclass{article}
\usepackage{spconf,amsmath,graphicx}
\usepackage{amsfonts}
\usepackage{xspace}
\usepackage{forloop}
\usepackage{CJK}
\usepackage{booktabs}
\usepackage{xfrac}
\usepackage{multicol,multirow}
\usepackage{color,xcolor}
\usepackage{siunitx}
\usepackage[pagebackref,breaklinks,colorlinks]{hyperref}
\usepackage{pict2e,picture} 
\usepackage{tikz}   
\usepackage{pifont} 

\newsavebox\CBox
\newlength\CLength
\def\numcircledpict#1{\sbox\CBox{#1}%
  \ifdim\wd\CBox>\ht\CBox \CLength=\wd\CBox\else\CLength=\ht\CBox\fi
    \makebox[1.5\CLength]{\makebox(0,1.5\CLength){\put(0,0){\circle{1.5\CLength}}}%
    \makebox(0,1.5\CLength){\put(-.5\wd\CBox,0){#1}}}}

\usepackage[capitalize]{cleveref}
\crefname{section}{Sec.}{Secs.}
\Crefname{section}{Section}{Sections}
\Crefname{table}{Table}{Tables}
\crefname{table}{Tab.}{Tabs.}

\def\mathbi#1{\textbf{\em #1}} 
\newcommand{\defcal}[1]{\expandafter\newcommand\csname 
	c#1\endcsname{{\mathcal{#1}}}}
\newcommand{\defbb}[1]{\expandafter\newcommand\csname 
	b#1\endcsname{{\mathbb{#1}}}}
\newcommand{\defbf}[1]{\expandafter\newcommand\csname 
	bf#1\endcsname{{\mathbf{#1}}}}
\newcommand{\defbk}[1]{\expandafter\newcommand\csname 
	bk#1\endcsname{{\mathfrak{#1}}}}
\newcommand{\defbi}[1]{\expandafter\newcommand\csname 
	bi#1\endcsname{{\mathbi{#1}}}}
\newcounter{calBbCounter}
\forLoop{1}{26}{calBbCounter}{
	\edef\letter{\Alph{calBbCounter}}
	\expandafter\defcal\letter
	\expandafter\defbb\letter
	\expandafter\defbf\letter
	\expandafter\defbk\letter
	\expandafter\defbi\letter
}

\makeatletter
\DeclareRobustCommand\onedot{\futurelet\@let@token\@onedot}
\def\@onedot{\ifx\@let@token.\else.\null\fi\xspace}

\def\eg{\emph{e.g}\onedot}

\def\etc{\emph{etc}\onedot}

\title{Visual-Aware Text-to-Speech\thanks{$*$~This work was done at JD Explore Academy.}}

\name{Mohan Zhou$^{1\dagger}$, Yalong Bai$^{2\dagger}$\thanks{$\dagger$~Equal contribution.\quad$\S$ Corresponding author.}, Wei Zhang$^2$, Ting Yao$^2$, Tiejun Zhao$^{1\S}$, Tao Mei$^2$}
\address{$^1$Harbin Institute of Technology, Harbin, China\quad$^2$JD Explore Academy, Beijing, China}

\begin{document}
%
\maketitle

\begin{abstract}
Dynamically synthesizing talking speech that actively responds to a listening head is critical during the face-to-face interaction. For example, the speaker could take advantage of the listener's facial expression to adjust the tones, stressed syllables, or pauses. In this work, we present a new visual-aware text-to-speech (VA-TTS) task to synthesize speech conditioned on both textual inputs and sequential visual feedback (e.g., nod, smile) of the listener in face-to-face communication. Different from traditional text-to-speech, VA-TTS highlights the impact of visual modality. On this newly-minted task, we devise a baseline model to fuse phoneme linguistic information and listener visual signals for speech synthesis. Extensive experiments on multimodal conversation dataset ViCo-X verify our proposal for generating more natural audio with scenario-appropriate rhythm and prosody.
\end{abstract}
\section{Introduction} \label{sec:intro}
Nowadays, the real-time face-to-face interaction between the customer and digital agent has attracted extensive attention due to its great potential for Metaverse and the next generation of human-computer interaction, e.g., digital humans, virtual agents, and social robots. By linking the natural language dialog backend and the audio-conditioned conversational video generator, the Text-to-Speech (TTS) system plays an essential role in converging the intelligence and expressiveness of digital agents. 

New scenarios lead to new technology demands. For face-to-face communication, the speaker verbally transmits information to the listener, and the listener responds to the speaker in real-time through non-verbal behaviors, e.g., affirmative nod, smiling, and head shake. Such two-way collaborative behaviors circulate the whole procedure, i.e., when the speaker is talking, they simultaneously perceive listeners' feedback and adjust their follow-up speech styles. From an acoustic standpoint, the fundamental frequency, intensity, or durational variations will adapt accordingly along with the procedure. Following social psychology and anthropology, these changes are function-specific and could be well-traced. For instance, listeners' expressions of confusion may evoke speakers' small pauses or longer duration. Similarly, agree/happy signals would excite speakers' passion and then lead to a higher fundamental frequency, larger intensity, and shorter duration. In view that classical TTS synthesizes speech from textual inputs only and the process is solely controlled by the speaker, it is impractical to directly apply the classical TTS for this communication scenario. To respond realistically to human perception in communication, synthesized speech should not only contain natural-sounding vocal inflection, rhythm, and prosodic features, but also necessitate cooperation with the listener.

\begin{figure}[t]
    \centering
    \includegraphics[width=0.9\linewidth]{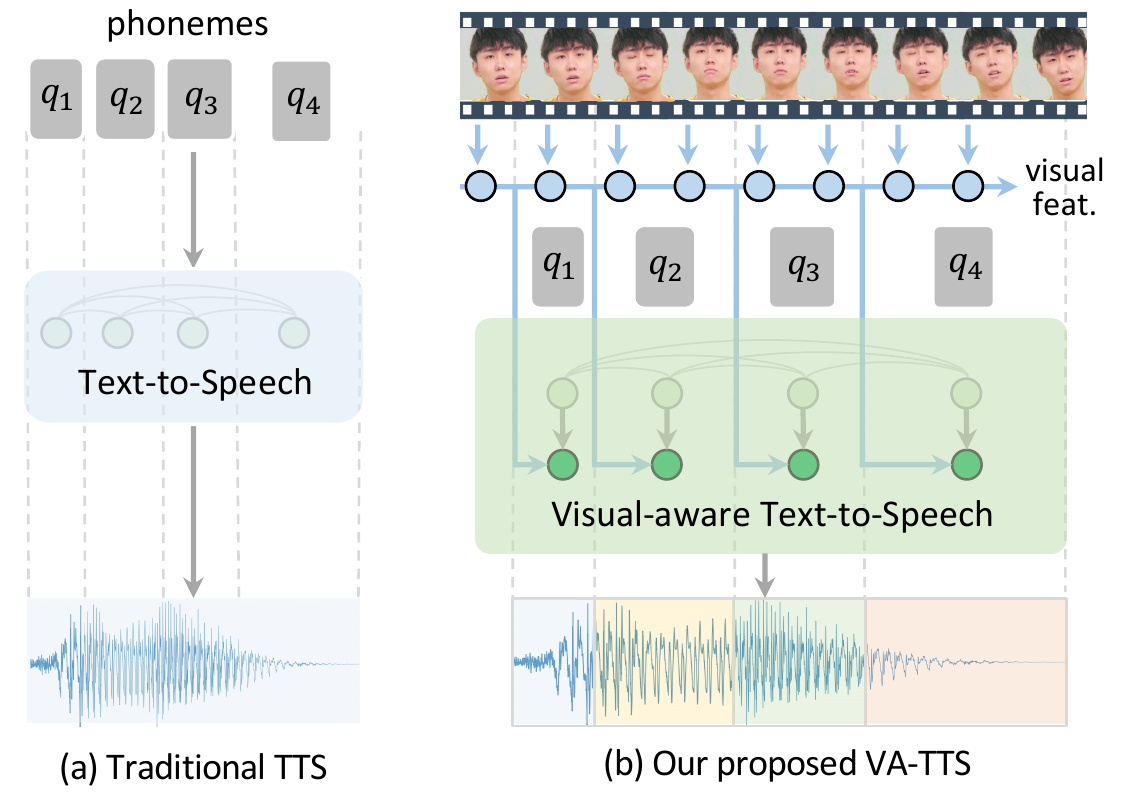}
    \caption{Illustrations of the traditional TTS and our proposed VA-TTS with extra listener video frames. Different colors in the waveform denote the prosody refinement responding to the listener's feedback from the previous moment.}
    \label{fig:arch}
\end{figure}

\begin{figure*}[!ht]
    \centering
    \includegraphics[width=0.96\linewidth]{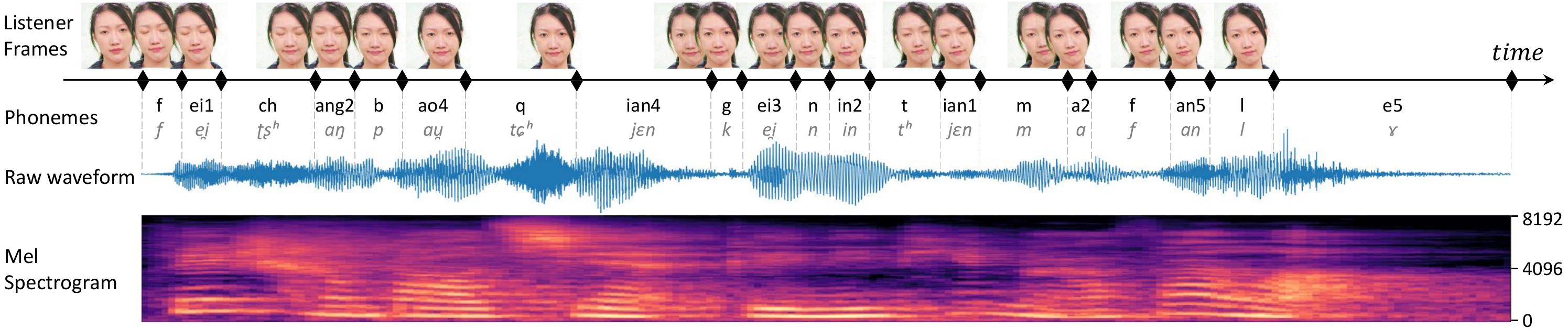}
    \caption{An example result of phoneme-listener alignment on ViCo-X dataset. Both the standard Chinese phonemes and IPA are annotated. We visualize the listener frames \textit{before} each phoneme starts and plot the mel spectrogram in log-scale.}
    \label{fig:phoneme_listener_align}
\end{figure*}

To alleviate this issue, we propose a new task, namely Visual-Aware Text-to-Speech (VA-TTS), to incorporate the visual feedback signals of listeners during text-conditioned speech synthesis. Different from the traditional TTS, VA-TTS synthesizes voice in a sequential manner by dynamically modifying the properties of speech according to the listener's behaviors in real-time, as shown in~\cref{fig:arch}. In this way, the generated speech is expected to be sensitive to dialog scenario changes and self-adaptive to scenario-appropriate rhythm and prosodic. The new task introduces ``interactivity'' to Text-to-Speech and would be critical to intelligent human-computer interaction and a wide range of user-friendly applications.

Furthermore, we introduce a simple but efficient VA-TTS baseline method. By training on the high-quality multi-modal multi-speaker Mandarin speech dataset ViCo-X~\cite{vicox}, our method would refine the output of traditional TTS at the phoneme level and predict more appropriate fundamental frequency, intensity, and pitch by fusing the video features and the semantic/acoustic information of audio. The proposed method is shown capable of generating more realistic speech by the quantitative results in~\cref{sec:experiments}. 
\section{Task Description} \label{sec:overview}
Following a text processing font-end, which transforms the textual inputs into phonemes $\cQ=\{q_1, q_2, \cdots, q_n\}$, a common neural network-based TTS system $\cA=\bfG(\cQ)$ consists of two main components: acoustic model, and vocoder. The acoustic model generates acoustic features from $\cQ$, which are always represented as line spectral pairs~\cite{itakura1975line}, linear-spectrograms~\cite{wang2017tacotron} or mel-spectograms~\cite{li2019neural, ren2020fastspeech}, \etc. The vocoder processes the acoustic features and generates the final waveform $\cA$. Such a pipeline can simulate human recordings and works well for presentation-oriented applications. For a specific speaker with given phonemes inputs $\cQ$, the waveform $\cA$ synthesized by a well-trained $\bfG$ is unique and independent from any other external factors.

However, in face-to-face interactive conversation scenarios, the non-verbal feedback from listeners can heavily influence the speaker's speech prosody reflecting as adaptively modifications of pitch, energy, and duration for the predetermined $\cQ$. As shown in~\cref{fig:phoneme_listener_align}, the voice of phonemes \textit{fei1chang2} tends to be low energy with negative feedback, and the duration of phonemes \textit{ao4qian4} become longer than other phonemes during the listener gazing. Obviously, both the semantic information from textual inputs and the real-time emotional feedback are essential to scenario-appropriate speech synthesis.

In this paper, we propose the Visual-Aware Text-to-Speech (VA-TTS) facing face-to-face interactive conversation scenarios. In addition to original input phonemes $\cQ$, we introduce the listener frames $\cV=\{v_1, v_2, \cdots, v_m\}$ to speech synthesis task. As shown in~\cref{fig:arch}, the VA-TTS need to perceive the listeners' information and understand their non-verbal visual signals, \textit{e.g.} head motion, facial expression, \etc, and then synthesize audio $\cA'=\bfG'_A(\cQ, \cV)$ with refined communication-oriented prosody. Same with traditional TTS, the textual inputs or phonemes can still be formed as parallel inputs. However, respecting practical applications, the visual frames $\cV$ should be treated as sequential inputs. Generally, the VA-TTS models accept phoneme inputs at once and then synthesize speech by adaptively adjusting prosody to the sequential listener's visual feedback.

\section{Baseline System} \label{sec:method}
In compliance with the VA-TTS task definition, we also introduce a simple but efficient VA-TTS baseline system consisting of three main components: training data preprocessing, listener feedback representation, and visual-aware prosody modeling. 

\subsection{Training Data Preprocessing} \label{sec:data_desc}
We apply the multimodal conversational dataset ViCo-X~\cite{vicox} for VA-TTS model training and evaluation. Compared to the existing TTS datasets (\eg, LJSpeech~\cite{ljspeech17}, VCTK~\cite{vctk}, \etc), ViCo-X possesses a unique visual modality that maintains the video frames for both the speaker and listener, which makes it possible to model the impacts from listener to audio. This dataset is constructed based on real scenario E-commerce conversation corpus ~\cite{chen2019jddc}. There are 10 speakers in the dataset. All audios are in Mandarin and recorded by high-fidelity microphones (\SI{48}{\kHz}, \SI{16}{\bit} depth).

In ViCo-X, each recording contains accompanying transcriptions in Chinese characters. To alleviate the mispronunciation problem, we convert the text sequence into phoneme sequence with an open-source Mandarin grapheme-to-phoneme converter\footnote{\url{https://github.com/GitYCC/g2pW}}. Then, to align listener frames to phoneme level audio inputs, instead of extracting the phoneme duration using a pre-trained auto-regressive TTS model, we use Montreal forced alignment (MFA)~\cite{mcauliffe2017montreal} tool to extract the phoneme duration. 

For the $i$-th phoneme $q_i\in\cQ$, we can locate its starting frame $v_{\hat{a}_i}\in\cV$. Considering the phoneme is the atomic unit in TTS, to simplify this task, a viable prerequisite setup is that the ongoing listener frames $\{v_{\hat{a}_i}...v_{\hat{a}_{i+1}-1}\}$ during the current phoneme duration are not used for modeling prosody of $q_i$. Moreover, assuming the time interval between two adjacent video frames is $\tau$, and the audio synthesis time cost of our method is $\mathcal{T}$ for each phoneme. To ensure the coherence of the generated speech, the sequential inputs should be fixed $\varphi$ steps in advance, where $\varphi$ is the smallest possible integer that satisfies the condition of $\varphi\tau\ge\mathcal{T}$. In other words, the prosody of $q_i$ is conditioned by all phonemes $\cQ$ and the past listener's video frame sequence:
\begin{equation}
\cV_{a_i}=\{v_1,...,v_{a_i}\}, a_i=\hat{a}_i-\varphi.
\end{equation}
Since all videos in ViCo-X are 30 FPS, and the $\mathcal{T}$ of our baseline method can geneate prosodic features in about \textbf{\SI{2.67}{\ms}}, we set $\varphi$ as 1 for all experiments in this paper. An example result of phoneme-listener alignment is shown in~\cref{fig:phoneme_listener_align}. 

\subsection{Listener Feedback Representations}
We aim to extract two representative features from the listener head frames: \textit{head motions}, and \textit{facial expressions}. Thanks to the state-of-the-art deep learning-based 3D face reconstruction model~\cite{deng2019accurate}, we can get the decoupled facial 3DMM~\cite{blanz1999morphable} coefficients easily. Specially, for listener head in each video frame, we can get the head reconstruction coefficients $\{\alpha, \beta, \delta, p, \gamma\}$ denoting the identity, expression, texture~\cite{cao2013facewarehouse,bfm09}, pose and lighting~\cite{ramamoorthi2001efficient}, respectively.  In order to drop the redundancy pixel and irrelevant identity information, we only keep the dynamic and identity-independent features $m=(\beta, p)$ as the listener feedback representations. Together, the head motions and facial expressions before the $i$-th phoneme can be tracked from $\cV_{a_i}$ as feature vector sequence $\cM_{a_i}=\{m_1, \cdots, m_{a_i}\}$.

To enrich the temporal information of listener feedback, a multi-layer recurrent encoder module $\bfE_v$ is applied on $\cM_{a_i}$, outputting the hidden states $\{h_1, \cdots, h_{a_i}\}$ of each time stamp in sequence as 
\begin{equation}
h_{a_i}=\bfE_v(h_{a_i-1}, m_{a_i}).
\end{equation}
Noting that in our scenario, the VA-TTS model is expected to receive the streaming input of listener frames where future information is not available. The listener feedback representations can be further aligned to the input phonemes $\cQ=\{q_1, \cdots, q_n\}$ as $\cH = \{h_{a_1}, \cdots, h_{a_n}\}$. 

\subsection{Visual-aware Prosody Modeling}
Inspired by~\cite{sorin2015coherent, rathina2012basic}, we use the fused feature of pitch $e^p_i$, energy $e^i_i$, and duration $e^d_i$ to represent the prosody $p_i=\{e^p_i, e^i_i, e^d_i\}$. For phonemes input $\cQ$, we can first get the classical TTS synthesized voice $\cA$, and then extract the contextualized speech representations $s_i$ for the corresponding synthesized phoneme $q_i$. Intuitively, the prosody for each phoneme during conversation should be determined by speaker identity $I$, audio characteristics (about voice synthesized by the classical TTS for the envisaged general scenario) $s_i$, phoneme $q_i$, and the listener features $h_{a_i}$ jointly. Thus, we apply stacked multi-head self-attention (MHSA) blocks for captures the key correlations among these features from different modality and outputs the fused feature:
\begin{equation}
\begin{aligned}
    x_i^{(0)} &= f(I; s_i; q_i; h_{a_i}; g^{(0)}_i)\\
    y_i^{(k)} &= x_i^{(k-1)} + \mathrm{MHSA}(\mathrm{LN}(x_i^{(k-1)})) \\
    x_i^{(k)} &= y_i^{(k)} + \mathrm{FFN}(\mathrm{LN}(y_i^{(k)})) \\
\end{aligned}
\end{equation}
where operation $\parallel$ joins two tensors over channel dimension, $g_i$ is the feature of a special symbol \texttt{[FUSION]} token added in front of input feature of MHSA, $k$ denotes the amount of the stacked MHSA blocks, and $f$ is non-linear feature transformation for projecting features from different modalities into a uniform scale. Then, we can infer the expected prosody feature $\bar{p}_i=\{\bar{e}^p_i,\bar{e}^i_i,\bar{e}^d_i\}$
from the composition representations $g^{(k)}_i)$ by a fully-connected neural layer.

Given the prosody representation $\hat{p}_i$ extracted from the ground-truth conversation speech by using an open-source audio processing libary~\cite{mcfee2015librosa} (we use the pitch spectrogram instead of pitch contour to avoid the high variations~\cite{ren2020fastspeech}), the whole model can be optimized by minimizing the $L_2$ distance between the prosody refinement $\bar{p}_i$ and the ground-truth $\hat{p}_i$.

The last step to get visual-aware text-to-speech results is Prosody Transfer, which has been well-studied in many recent related works. Here, to highlight the listener feedback that can help the prosody recovery of the traditional TTS system, we just feed the refined prosody $\bar{\cP}=\{\bar{p}_1,\cdots,\bar{p}_n\}$ to FastSpeech2 framework to generate final audio $\cA'$.

\textbf{Inference}\quad In a practical dialogue application scenario, the input to VA-TTS system consists of textual inputs determined from an intelligent dialogue backend (pre-processed as $\cQ$) and an online listener video stream $\cV$. The main difference with traditional TTS is that the VS-TTS model needs to capture the video features corresponding to the current phoneme $p_i$ pronunciation, which can be pre-located according to the accumulation of predicted durations of the previous phonemes as $a_i=\sum_{k=1}^{i-1}\bar{e}^d_{k}-\varphi$. After that, the output audio can be sequentially refined by the predicted prosody. 

\section{Experiments} \label{sec:experiments}
The proposed baseline system is evaluated on the ViCo-X multi-speaker conversational audio-video dataset. For listener frames, we extract the 3DMM coefficients~\cite{deng2019accurate} following with the commonly used toolkit\footnote{\url{https://github.com/microsoft/Deep3DFaceReconstruction}}, therefore, we can get the relative dynamic, identity-independent features $m=(\beta, p)\in\mathbb{R}^{1\times 70}$ for each image. 

We finetune the pretrained FastSpeech2~\cite{ren2020fastspeech} model on $\cD_{train}$ and use the pretrained HiFi-GAN as a vocoder to get baseline results. For FastSpeech2 finetuning, the encoder part is frozen, and only the variance adaptor and mel-spectrogram decoder are updated. We still optimize with the Adam algorithm while using a smaller learning rate \SI{1e-3}{} to keep the finetuning process more stable.

For VA-TTS, there are multiple choices to implement the listener feedback sequence module $\bfG_m$, such as LSTM, GRU, or a Transformer encoder with a sliding window. Here we adopt LSTM for our baseline, since it has been widely used in many video sequence encoding tasks and can achieve stable state-of-the-art performance. Our VA-TTS model is trained with Adam optimizer with a learning of \SI{5e-4}{} decayed following cosine annealing rule), $\beta_1=0.9$ and $\beta_2=0.999$ for 240 epochs. The prosody features are optimized in log-scale in our experiments.

\subsection{Quantitative Analysis}
To quantify the quality and naturalness of synthesized audio samples with quantitative metrics, we apply several metrics used in early work to evaluate pairwise prosody similarity with the ground-truth audio, including the \textit{mel-cepstral distortion} (MCD$_{13}$)~\cite{mcd}, \textit{gross pitch error} (GPE)~\cite{gpe}, \textit{voice decision error} (VDE)~\cite{gpe}, and the \textit{F$_0$ frame error} (FFE)~\cite{ffe}. \cref{tab:mcd} shows reconstruction performance measured using the above metrics. Lower is better for all metrics.

\begin{table}[h]
    \centering
    \begin{tabular}{lcccc}
    \toprule
     & GPE & VDE & FFE & MCD$_{13}$ \\
    \midrule
     TTS & 20.23 & 9.54 & 20.90 & 6.11 \\
     VA-TTS  & 20.09 & 9.23 & 20.58 & 5.97\\
     \bottomrule
    \end{tabular}
    \caption{Objective prosody scores for TTS and VA-TTS. TTS denotes the FastSpeech2 model fine-tuned on ViCO-X.}
    \label{tab:mcd}
\end{table}

Moreover, to verify whether providing listener's visual feedback as input can indeed synthesize speech with more accurate pitch, energy, and duration, we compare the \textit{mean absolution error} (MAE) between the phoneme-wise pitch/energy/duration extracted from the generated waveform and the ground-truth speech. Ignoring the minor artifacts on short timescales, phonemes whose duration error is lower than 50\textit{ms} are not excluded. The results are shown in~\cref{tab:quantitative_results}. All these experimental results demonstrated that, compared with the traditional TTS, VA-TTS could synthesize speech audio with more similar prosodies to the ground-truth audio under face-to-face conversation scenarios.

\begin{table}[h]
    \centering
    \begin{tabular}{lccc}
    \toprule
        $\Delta$ & Pitch & Energy & Duration\\
    \midrule
         TTS & 37.02 &  5.32  & 143.73 \\
         VS-TTS & 35.60 & 5.08 & 135.97 \\
    \bottomrule
    \end{tabular}
    \caption{The averaged mean absolution error $\Delta$ of prosody characteristic between ground-truth audio and the synthesized audio $\cA$ from TTS / the refined audio $\cA'$ from VS-TTS. }
    \label{tab:quantitative_results}
\end{table}

\subsection{Qualitative Analysis}
\begin{figure}[h]
    \centering
    \includegraphics[width=\linewidth]{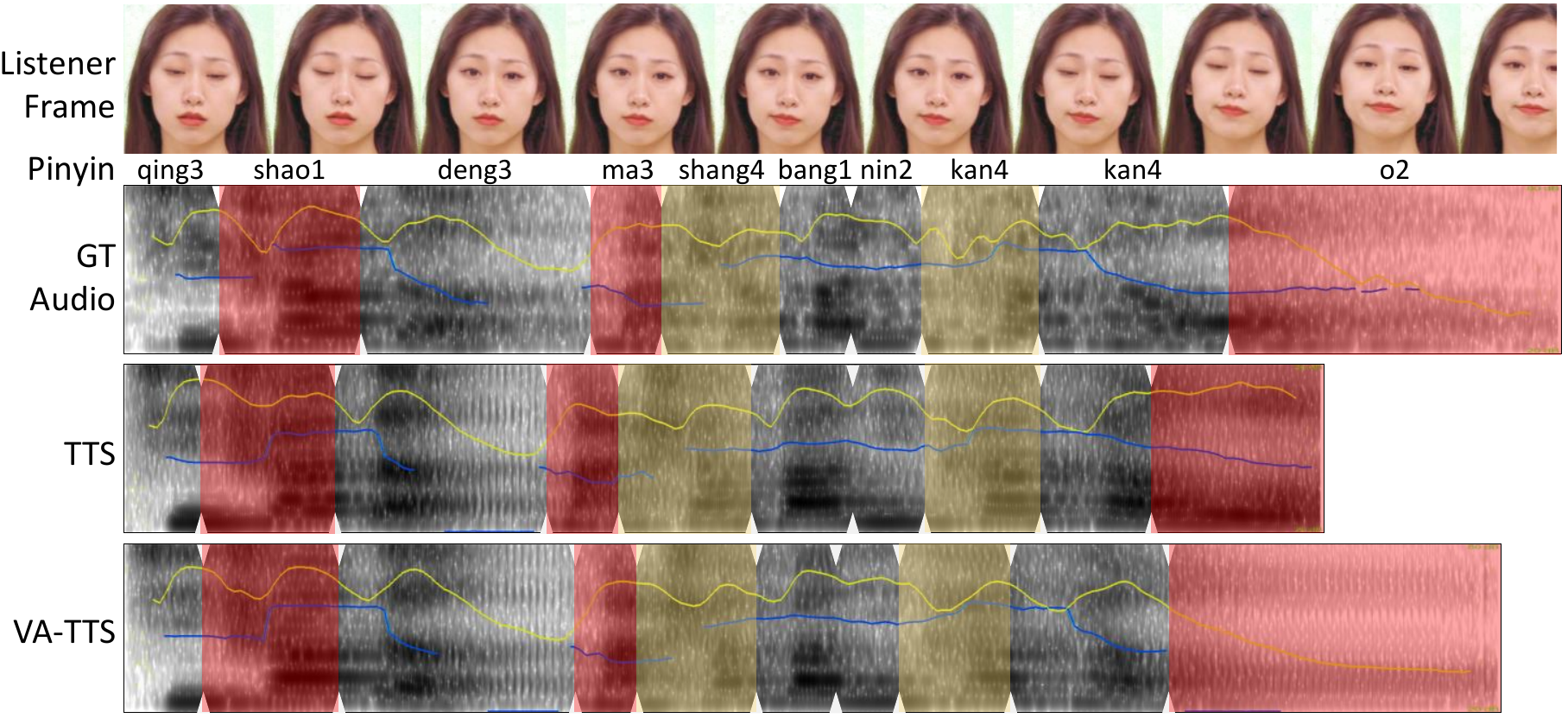}
    \caption{The mel-spectrograms of the voice. \textcolor{yellow}{Yellow} and \textcolor{blue}{blue} lines denote the energy and pitch respectively. We use Chinese Pinyin instead of phonemes to visualize the alignment for avoiding too much fragments. 
    }
    \label{fig:quantitive_results}
\end{figure}

We visualize the mel-spectorgrams as well as prosody features of the ground truth audio, traditional TTS generated audio and our VA-TTS refined audio in~\cref{fig:quantitive_results}. The corresponding text of exampel is ``\emph{Please wait a moment to help you solve it}''. We can observe that the VA-TTS demonstrate its ability to generate higher-quality audio samples by pitch correction (red blocks), energy finetuing (yellow blocks), and duration scaling, making existed TTS system more suitable for face-to-face communication scenraios. Meanwhile, some interesting patterns reveal how listener affect the speaker's audio, \eg, duration scaling becomes conspicuous when perceive listeners' negative feedbacks (last two pinyins).

\section{Conclusion} \label{sec:conclusion}
In this paper, we have proposed a new task, Visual-Aware Text-to-Speech, the first speech synthesizer utilizing the listener role in communication. Alongside this task, we also provide a baseline that aims to generate communication-satisfied prosody features and validate on ViCo-X dataset. Quantitative analysis and subjective evaluations are conducted to prove that our VA-TTS can synthesize interactable speech that more consistent with human perception. 

\textbf{Acknowledgement} This work was supported by the National Key R\&D Program of China under Grant No. 2020AAA0108600.

\bibliographystyle{IEEEbib}
\bibliography{strings,refs}

\end{document}